\documentclass[conference]{IEEEtran}
\IEEEoverridecommandlockouts
\usepackage{cite}
\usepackage{amsmath,amssymb,amsfonts}
\usepackage{algorithmic}
\usepackage{graphicx}
\usepackage{subfigure}
\usepackage{graphicx}
\usepackage{url}
\usepackage{tikz}
\usepackage{subfig}
\usepackage{multirow}
\usepackage{textcomp}
\usepackage{xcolor}
\usepackage{lipsum}
\def\BibTeX{{\rm B\kern-.05em{\sc i\kern-.025em b}\kern-.08em
    T\kern-.1667em\lower.7ex\hbox{E}\kern-.125emX}}

\newcommand\copyrightnotice{
    \begin{tikzpicture}[remember picture,overlay]
        \node[anchor=south,yshift=10pt] at (current page.south) {\fbox{\parbox{\dimexpr\textwidth-\fboxsep-\fboxrule\relax}{\copyrighttext}}};
    \end{tikzpicture}
}

\newcommand\copyrighttext{
  \footnotesize \textcopyright 2012 IEEE. Personal use of this material is permitted.
  Permission from IEEE must be obtained for all other uses, in any current or future
  media, including reprinting/republishing this material for advertising or promotional
  purposes, creating new collective works, for resale or redistribution to servers or
  lists, or reuse of any copyrighted component of this work in other works.
  DOI: \url{https://doi.org/10.1109/SIPAIM56729.2023.10373517}
}

\begin{document}

\title{Self-calibrated convolution towards glioma segmentation}

\author{
\IEEEauthorblockN{Felipe C. R. Salvagnini\IEEEauthorrefmark{1}, Gerson O. Barbosa\IEEEauthorrefmark{2},
Alexandre X. Falcão\IEEEauthorrefmark{3},
Cid A. N. Santos\IEEEauthorrefmark{4}
\IEEEauthorblockA{\IEEEauthorrefmark{1}\IEEEauthorrefmark{2}\IEEEauthorrefmark{4}\textit{Computational Photography Department (DFC) - Eldorado Institute - Campinas, Brazil}}
\IEEEauthorblockA{\IEEEauthorrefmark{1}\IEEEauthorrefmark{3}\textit{State University of Campinas (UNICAMP) - Campinas, Brazil}}
\IEEEauthorblockA{\IEEEauthorrefmark{2}\textit{São Paulo State University (UNESP) - Guaratinguetá, Brazil}\\
Email: \IEEEauthorrefmark{1}felip.eng@outlook.com.br 
\IEEEauthorrefmark{2}gerson.barbosa@unesp.br 
\IEEEauthorrefmark{3}afalcao@ic.unicamp.br} 
\IEEEauthorrefmark{4}cid.santos@eldorado.org.br}
}
\maketitle
\copyrightnotice

\begin{abstract}

Accurate brain tumor segmentation in the early stages of the disease is crucial for the treatment's effectiveness, avoiding exhaustive visual inspection of a qualified specialist on 3D MR brain images of multiple protocols (e.g., T1, T2, T2-FLAIR, T1-Gd). Several networks exist for Glioma segmentation, being nnU-Net one of the best. In this work, we evaluate self-calibrated convolutions in different parts of the nnU-Net network to demonstrate that self-calibrated modules in skip connections can significantly improve the enhanced-tumor and tumor-core segmentation accuracy while preserving the whole-tumor segmentation accuracy.

\end{abstract}


\section{Introduction}
\label{sec:introduction}

Two primary classes of cells compose the nervous system: \textit{neurons} and \textit{glia}. Each of them has specific functions. \textit{Neurons}, electrically excited cells, are responsible for transmitting electrical and chemical signals, enabling neuronal synapses \cite{pereda2014electrical}. While \textit{glial} cells guarantee the proper working of neurons by interacting with neuronal synapses and capillary networks and enabling blood-brain barriers \cite{fundamental_neuroscience_4}.

We can further divide \textit{glial} cells into three major types: \textit{microglia}, \textit{astrocytes}, and \textit{oligodendrocytes}. According to gene profile analysis and genetic modeling in mice, there is evidence that Glioblastomas (GBM) --- a grade IV glioma --- derive from those cells \cite{gbm_origin}. GBM is a particularly aggressive and malignant tumor and the most common malignant brain tumor in adults. 

Brain tumors are known for their high mortality and the complexity of treatment. In general, these two aspects relate to its location. Glioma, a type of primary brain tumor, is one of the most common cases \cite{louis2016cimpact}. Being almost 30\% of all primary brain tumors and 80\% of all malignant ones, they are responsible for most deaths caused by primary brain tumors \cite{weller2015glioma}.

Therefore, for the treatment of GBM cases, the problem of delineating tumor regions is of significant importance. Precisely identifying tumor tissues guides medical procedures. The absence of an automated solution forces radiologists to visually inspect 3D MR brain images of multiple protocols (e.g., T1, T2, T2-FLAIR, T1-Gd) -- an exhaustive and error-prone task. 

Active research on Convolutional Neural Networks (CNNs), as U-shaped encoder-decoder architectures, have been pursued for brain tumor segmentation~\cite{unet, unet_3d, nnunet}. Such semantic segmentation (SS) models require voxel-level annotation of many images for training and should identify relevant tumor regions for guiding the surgical procedure. Accurate voxel-level image annotation is challenging, which makes the available datasets (e.g., \textit{BraTs}~\cite{brats2021}) valuable to support research in this area. 

Nevertheless, most works focus on developing U-shaped networks employing plain 3D Convolutions. Despite the convolution's power to extract semantic features, more efforts must be made toward improving feature learning. This study sheds light on integrating self-calibrated convolutions in U-Shaped networks. We evaluate such modules in different segments of the nnU-Net network and conclude that self-calibration in skip connections is the best choice.

\section{Related Work}
\label{sec:related_work}

Towards GBM problems, Multi-modal Brain Tumor Image Segmentation Benchmark (\textit{BraTS}) is the standard benchmark. The \textit{BRATS} dataset was built in 2012 with the MICCAI (Medical Image Computing and Computer Assisted Intervention) conference. Initially, the dataset had 30 multi-contrast MR scans, but the latter version (2023) has 2040 \cite{brats2021, brats_2015}. Medical experts annotate tumor tissues into three classes: Gd-enhancing tumor (ET), peritumoral edematous/invaded tissue (ED), and the necrotic tumor core (NCR).

The \textit{BraTS} dataset targets heterogeneous image quality. Hence, brain tumor mpMRI (multi-institutional multi-parametric pre-operative magnetic resonance imaging) reflects many clinical practices across different institutions \cite{brats2021}. A review of past years' \textit{BraTS} leader-board --- winning methods --- shows a tendency to deep learning (since 2014) \cite{brats_2021_winner}. In the following years, methods introduced architectural improvements toward U-Shaped architectures (initially proposed by U-Net\cite{unet}) \cite{nnunet}.

Kamnitsas et al. won the 2017 competition through an Ensemble of Multiple Models and Architecture (EMMA) \cite{brats_2017_winner}. The 2018 winning solution proposed improvements such as an asymmetrical U-Net and an auto-encoder to carry out regularization \cite{brats_2018_winner}. In the following year, the winners employed a coarse-to-fine strategy \cite{brats_2019_winner}. Through a cascade network, they first predict a rough segmentation map, which a more complex U-Net further improves.

Lately, the nnU-Net (No-new-U-Net) won the \textit{BraTS} 2020 competition \cite{nnunet}. nnU-Net is a fully automatic framework for SS methods configuration. A set of heuristics guide the design of the U-Net architecture and pre- and post-processing operations. Additionally, their results demonstrate the effect of data augmentation on the model's performance. In 2021, authors extended nnU-Net exploring larger decoders, different normalization, and attention layers.

Across all winning solutions, from cascade and auto-encoder to attention layers, all U-shaped models employ, as building blocks, plain convolutional blocks (3D). The network encoder extracts semantic information from different resolutions, but there is room for improvement. Furthermore, winning approaches do not apply operations on skip connections. Hence, we explore this gap by investigating a module introduced to improve convolution by enabling long-range spatial and inter-channel dependencies around each spatial location \cite{self_calibrated}. Interestingly, this module, named self-calibrated convolutions (SC-Conv), improves the basic convolutional feature transformation process without modifying model architectures. Hence, our approach modifies the nnU-Net baseline by replacing plain convolutions with SC-Conv.

\section{Adding SC-Conv into U-shaped Networks}
\label{sec:SC}

Sequences of convolutional blocks aim to improve data representation for classification and segmentation tasks. An SC-Conv module contains five convolutional blocks~\cite{self_calibrated} to improve data representation further. Such modules may be used in sequence just as plain convolutional blocks are used. We explore the idea in~\cite{self_calibrated} to incorporate 3D SC-Conv modules into U-shaped networks (i.e., nn-UNet 3D low-resolution without five-fold ensemble).

\begin{figure}[h]
    \centering
    \includegraphics[width=0.5\textwidth]{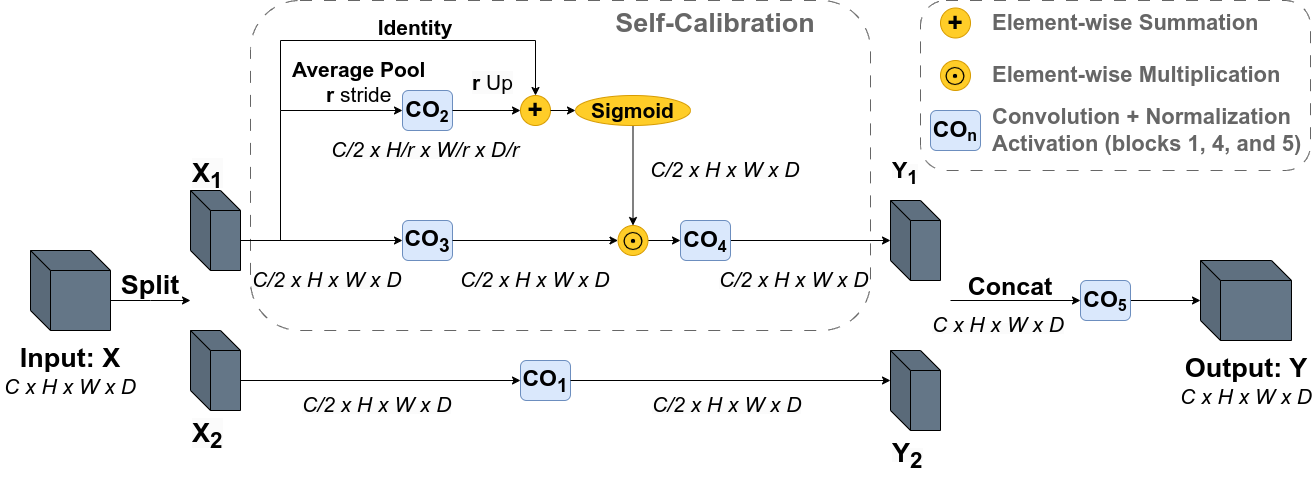}
    \caption{3D SC-Conv Module \cite{self_calibrated}.}
    \label{fig:sc_explanation}
\end{figure}

Figure \ref{fig:sc_explanation} illustrates an SC-Conv module (3D). First, the input image $X$ with $C$ channels is split into two images, $X_1$ and $X_2$, of $\frac{C}{2}$ channels each. A plain convolution layer performs such a split operation (i.e., convolution groups). Self-calibration is executed over $X_1$, downsampling it by average pooling with stride $r$ to pass through convolution and batch normalization ($CO_2$). The result is upsampled by interpolation with factor $r$ for element-wise summation with  $X_1$ followed by a sigmoidal activation, forcing an output within $[0,1]$. For self-calibration, this output is element-wise multiplied by the result of passing $X_1$ through another convolution and batch normalization block ($CO_3$). The result of self-calibration then passes through another convolution, batch normalization, and ReLU activation ($CO_4$), generating $Y_1$. In parallel, $X_2$ passes through another convolution, batch normalization, and ReLU activation ($CO_1$), generating $Y_2$. After concatenating $Y_1$ and $Y2$, the resulting image passes through another convolution, batch normalization, and ReLU activation ($CO_5$) to obtain the output $Y$. 

SC-Conv modules aim to encode semantics without introducing complexity overhead or hyper-parameters tunning. It has the following advantages. (1) It can model inter-channel dependencies, enlarging the fields of view. As such, it encodes larger and more accurate discriminative regions. (2) Instead of collecting global contexts, SC only considers the context around each spatial location. Consequently, it avoids the contamination of information from irrelevant regions. (3) It encodes multiscale information, which is highly desirable for object detection tasks \cite{self_calibrated}.

As SC-Conv could easily replace plain convolutional layer: we adapted the original code\footnote{https://github.com/MCG-NKU/SCNet} to work with Volumetric Data (Batch x Channel x Height x Width x Depth); replaced batch normalization by instance normalization, and ReLU by leaky-Relu, as we use nnU-Net; and evaluated such modules in different segments of the nn-UNet network, on encoder-decoder (simultaneously), and skip connections. For the case of the encoder and decoder, we replaced the 3D convolutional blocks with 3D SC-Conv blocks. In the case of skip connections, we added the module for each encoder level before concatenating extracted features for decoding.

\section{Experimental setup}
\label{sec:method}

This section describes our experimental setup to evaluate the impact of adding self-calibrated convolutions into nn-UNet. We describe the dataset, our baseline (nnU-Net \cite{nnunet}), and its modifications. We also present the training procedure and evaluation metric.

\subsubsection{Dataset}

\begin{figure}[h]
    \centering
    \includegraphics[width=0.4\textwidth]{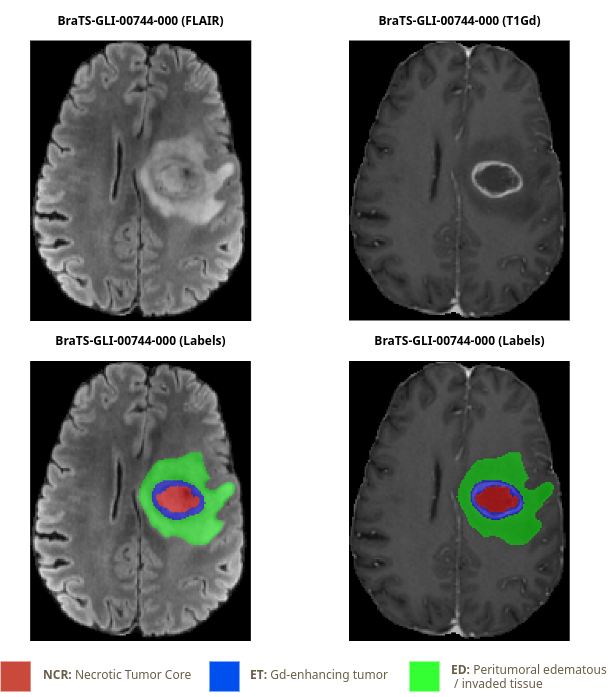}
    \caption{Data instance from BraTS23. One can visualize NCR (hypointense) and ET (hyperintense) tissues on T1Gd, and ED (hyperintense) on T2-FLAIR protocols.}
    \label{fig:glioma_sample}
\end{figure}

We employed data from the BraTS 2023 Adult Glioblastoma challenge as our training and validation data. The data are the same as the 2021 challenge \cite{brats2021}. Moreover, as labels are available for the competition training data, we split the training data (1251) into the train (80\%) and validation (20\%) data. Each training instance comprises 4 MRI modalities: 1. native (T1); 2. post-contrast T1 Weighted (T1Gd); 3. T2-weighted (T2); and 4. T2 Fluid Attenuated Inversion Recovery (T2-FLAIR). Finally, each training instance has a ground truth delineating the glioma into three classes, shown in image \ref{fig:glioma_sample}.

\subsubsection{Network Structure}

The BraTS leader-board for the last years shows a tendency to nnU-Net-based approaches. Hence, we choose our baseline using the out-of-the-box nnU-Net framework \footnote{https://github.com/MIC-DKFZ/nnUNet}. We first pre-processed the BraTS 2023 Glioblastoma dataset using the nnU-Net pre-processor for such an end. Then, the framework defines a 3D U-Net architecture for image patches (128x128x128), setting convolutional blocks, normalization operations, and deep supervision. Deep supervision allows the calculus of loss for the three last decoder's levels, helping gradient flow \cite{nnunet, nn_unet_nivida}.

Then, based on the baseline model defined by nnU-Net, we conducted three experiments:

\begin{enumerate}
\item[M1] 3D SC-Conv replaces all convolution blocks in the encoder and decoder;
\item[M2] 3D SC-Conv is used only for skip connections;
\item[M3] The combination of M1 and M2, with 3D SC-Conv in all parts, encoder, decoder, and skip connections.
\end{enumerate}

\begin{figure}
    \centering
    \includegraphics[width=0.4\textwidth]{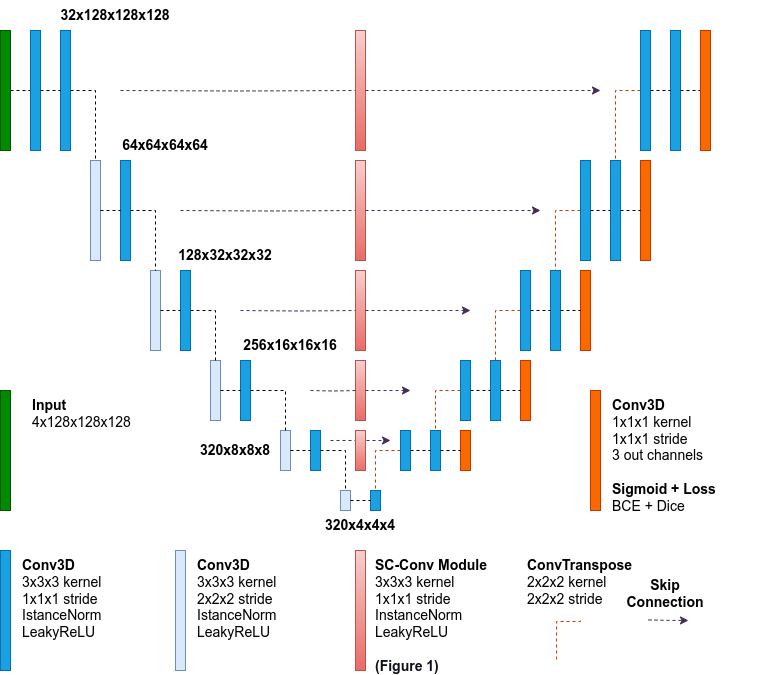}
    \caption{Best architecture found (nnU-Net\cite{nnunet} with SC-Conv on skip-connections).}
    \label{fig:arch}
\end{figure}

Figure \ref{fig:arch} shows our expansion of nnU-Net with SC-Conv (M2).

Finally, the only modifications made to the original code of SC-Conv are 3D Convolutions, leaky-Relu, and instance normalization, as suggested by nnU-Net.

\subsubsection{Training procedure}

We conducted all experiments in an NVIDIA A100 GPU (40Gb). The loss function --- a combination of binary cross entropy and Dice --- does not optimize the tumor classes directly, but the following: ET, Tumor Core (NCR $\cup$ ET), and Whole Tumor (NCR $\cup$ ET $\cup$ ED). To normalize our input images (ZScore normalization) and augment them, we kept the nnU-Net proposed operations across all experiments \cite{nnunet}. Despite working with image patches, the training uses a batch size of 2 to avoid the burden of GPU memory. This batch size also justifies using instance normalization \cite{instance_normalization}.

All experiments were executed for 260 epochs using a Poly learning rate scheduler, initially set as 0.01. As the optimizer, we employed SGD with Nesterov momentum ($\mu = 0.99$).

\subsubsection{Evaluation metric}

 We used Dice Score Coefficient (DSC) to measure the similarity between the segmentation masks (predicted vs ground truth).

\section{Results}
\label{sec:results}

As previously introduced, we provided a fair comparison for 3D SC-Conv against 3D Convolution, keeping all other network parameters across experiments. Table \ref{t:results} summarizes our results.

\begin{table}[!htbp]
\centering
\begin{tabular}{l|cccc}
\multicolumn{1}{c|}{\multirow{2}{*}{\textbf{Approach}}} & \multicolumn{4}{c}{\textbf{Dice (\%, $\uparrow$)}}                                                          \\ \cline{2-5} 
\multicolumn{1}{c|}{}                         & \multicolumn{1}{l|}{\textbf{ET}} & \multicolumn{1}{l|}{\textbf{TC}} & \multicolumn{1}{l|}{\textbf{WT}} & \textbf{AVG} \\ \hline

Baseline                                            & \multicolumn{1}{c|}{$83.49_{22}$}  & \multicolumn{1}{c|}{$90.19_{15}$}  & \multicolumn{1}{c|}{$\boldsymbol{91.53_{10}}$}  & $88.42_{14}$   \\

M1                                      & \multicolumn{1}{c|}{$83.08_{22}$}  & \multicolumn{1}{c|}{$89.68_{15}$}  & \multicolumn{1}{c|}{$91.13_{10}$}  & $87.97_{14}$   \\

M2                                 & \multicolumn{1}{c|}{$\boldsymbol{84.32_{22}}$}  & \multicolumn{1}{c|}{$\boldsymbol{90.46_{15}}$}  & \multicolumn{1}{c|}{$91.10_{10}$}  & $\boldsymbol{88.65_{14}}$   \\

M3                                 & \multicolumn{1}{c|}{$82.98_{23}$}  & \multicolumn{1}{c|}{$89.63_{16}$}  & \multicolumn{1}{c|}{$90.65_{10}$}  & $87.75_{14}$   \\ \hline
\end{tabular}
\caption{Dice score metric (Standard deviation as subscript).}
\label{t:results}
\end{table}

Results showed that adding 3D SC-Conv on the encoder and decoder degraded our performance. However, adding them on skip-connections helps to propagate high-quality features to the decoder, improving our performance. Furthermore, high standard deviation values make it difficult to conclude further improvements. However, this is explained by BraTS variability, as the dataset represents different institutions under standard clinical conditions but with diverse equipment and imaging protocols \cite{brats2021}.

\begin{figure}[]
    \centering
    \subfigure[]{\includegraphics[width=0.15\textwidth]{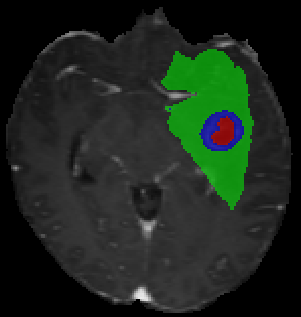}}
    \subfigure[]{\includegraphics[width=0.15\textwidth]{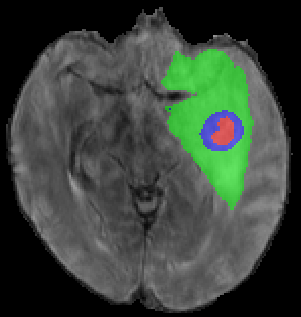}}
    \\
    \subfigure[]{\includegraphics[width=0.14\textwidth]{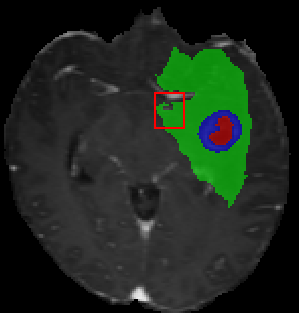}}
    \subfigure[]{\includegraphics[width=0.14\textwidth]{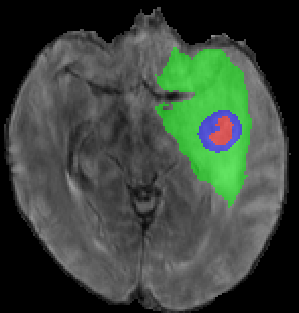}}
    \\
    \subfigure[]{\includegraphics[width=0.14\textwidth]{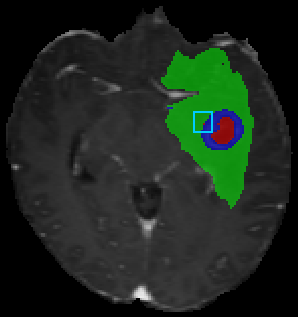}}
    \subfigure[]{\includegraphics[width=0.14\textwidth]{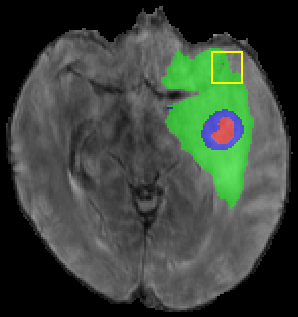}}
    \caption[Shorter figure caption]{Results for BraTS subject 00019. The first column shows T1-Gd images, and the second one, T2-Flair. a-b are results with SC-Conv on skip-connections, c-d the ground truth, and e-f the baseline's result.}
    \label{f:reslicing_2}
\end{figure}

Finally, Figure \ref{f:reslicing_2} shows an inference made by our model and the baseline. Shown on the left is the T1Gd image, and T2-FLAIR on the right. The first line shows our model's inference on both images, while the third line shows the baseline segmentation masks. On the center line is the ground truth.

Both models struggle to classify the peritumoral edema in the edge region correctly. Also, models estimate a convexity, but a cavity (red square) exists. Moreover, the baseline loses a small hole of edema. At the same time, our model correctly classifies it, depicting the potential of SC-Module to collect spatial location (yellow square). In the same context, the baseline also predicts necrotic tumor core in invalid positions. In contrast, a better context understanding prevents it (blue square). The same behavior is observed for other images.

\section{Conclusion}
\label{sec:conclusion}

Our experiments raise the following takeaways: As introduced by nnU-Net authors, aggressive augmentations substantially improve segmentation performance, while architecture design shows slight enhancement. Hence, future work may focus on designing more suited augmentations or network architectures to tackle data variance, a characteristic of medical data from multiple institutions; Secondly, SC-Conv on skip-connections improve performance, but further experiments are required: on skip-connection to validate if using only 3D convolutions also shows improvements; on encoder-decoder, exploring different expansion and pooling rates to explore more spatial features at U-Net higher level. Lastly, it also shows the potential of the nnU-Net to define a baseline and experiment with additional modifications easily.

\section*{Acknowledgment}
The authors thank \textit{Eldorado Institute} for the computational resources, and the third author thanks CNPq (Proc. 303808/2018-7) for the financial support.

\bibliographystyle{IEEEtran}  
\bibliography{ref}  

\end{document}